\documentclass[runningheads,a4paper]{sty/llncs2e/llncs}

\usepackage{style}
\usepackage{xcolor}
\usepackage{enumitem}
\usepackage{paralist}
\usepackage{textcomp}
\usepackage{tabularx}
\usepackage{dcolumn}
\usepackage{graphicx}
\usepackage{amssymb}
\usepackage{amsmath}

\newcolumntype{Z}{D..{1.3}}

\usepackage[USenglish]{babel}

\usepackage{tikz}
\usetikzlibrary{decorations.pathreplacing}
\usetikzlibrary{shadows}
\usetikzlibrary{calc}
\usetikzlibrary{shapes}
\usetikzlibrary{backgrounds}

\definecolor{green}{RGB}{0,170,0}
\definecolor{lightgreen}{RGB}{138,255,138}
\definecolor{lightblue}{RGB}{138,138,255}
\definecolor{lightorange}{RGB}{255,212,138}

\newcommand{\underbraced}[2][]{%
  $\sbox0{$\vcenter{\hbox{#2}}$}%
    \vphantom{\copy0}
  \kern-\nulldelimiterspace
  \underbrace{\vrule width0pt depth \dimexpr\dp0 + .3ex\relax\box0}_{\text{#1}}$}

\def\imgplot{plot coordinates{
  (0,-6)
  (1,-2)
  (2,-3)
  (3,-1)
  (4,-6)
  (5,-4)
  (6,-6)
  (7,-5)
  (8,-6)
};}%

\def\slantfrac#1#2{\kern.1em^{#1}\kern-.1em/\kern-.1em_{#2}}

\newif\ifanonimize

\anonimizefalse

\begin{document}
\mainmatter  

\title{Region segmentation for sparse decompositions: better brain
parcellations from rest fMRI}

\titlerunning{Automatic region extraction for atlases extracted from resting-state fMRI}

\ifanonimize
\else
\author{Alexandre \textsc{Abraham}\inst{1}\textsuperscript{,}\inst{2} \and
        Elvis \textsc{Dohmatob}\inst{1}\textsuperscript{,}\inst{2} \and
        Bertrand \textsc{Thirion}\inst{1}\textsuperscript{,}\inst{2} \and
        Dimitris \textsc{Samaras}\inst{3}\textsuperscript{,}\inst{4} \and
        Gael \textsc{Varoquaux}\inst{1}\textsuperscript{,}\inst{2}}
\authorrunning{Alexandre \textsc{Abraham} et al.} 
%
%
\institute{Parietal Team, INRIA Saclay-\^{I}le-de-France, Saclay, France\\
\email{alexandre.abraham@inria.fr}
\and
CEA, DSV, I\textsuperscript{2}BM, Neurospin b\^{a}t 145,
91191 Gif-Sur-Yvette, France
\and
Stony Brook University, NY 11794, USA
\and
Ecole Centrale, 92290 Ch\^atenay Malabry, France
}
\fi

\maketitle

\begin{abstract}

Functional Magnetic Resonance Images acquired during rest\-ing-state
provide information about the functional organization of the brain through
measuring correlations between brain areas. 
Independent components analysis is the reference approach to
estimate spatial components from weakly structured data such as brain
signal time courses; each of these components may be referred to as a
\textit{brain network} and the whole set of components can be
conceptualized as a \textit{brain functional atlas}.
Recently, new methods using a sparsity prior have emerged to deal with
low signal-to-noise ratio data. 
However, even when using sophisticated priors, the results may not be
very sparse and most often do not separate the spatial components into
brain regions.
This work presents post-processing techniques that automatically sparsify
brain maps and separate regions properly using geometric operations, and
compares these techniques according to faithfulness to data and stability metrics.
In particular, among threshold-based approaches, hysteresis thresholding
and random walker segmentation, the latter improves significantly the
stability of both dense and sparse models.

\end{abstract}

\section{Introduction}

Functional connectivity between brain networks observed during resting state
functional Magnetic Resonance Imaging (R-fMRI) is a promising source of
diagnostic biomarkers, as it can be measured on impaired subjects such as
stroke patients \cite{varoquaux2010b}. 
However, its quantification highly depends on the
choice of the brain atlas.
A brain atlas should be \textit{i)}
consistent with neuroscientific knowledge \textit{ii)} as faithful as
possible to the original data and \textit{iii)} robust to inter-subject
variability.

Publicly available atlases (such as structural
\cite{tzourio-mazoyer2002a} or functional \cite{Yeo2011} atlases) went
through a quality assessment process and are reliable. 
To extract a data driven atlas from R-fMRI, Independent Component Analysis (ICA) remains the reference
method. 
In particular, it yields some additional flexibility to adapt the number of
regions to the amount of information available.
Networks extracted by ICA are full-brain and require a
post-processing step to extract the salient features, i.e., brain regions, which is often done
manually \cite{kiviniemi2009} (see figure~\ref{fig:ica}). 
To avoid post-processing and directly extract
regions, more sophisticated approaches rely on sparse,
spatially-structured priors \cite{abraham2013}. 
Indeed, maps of functional networks or regions display a small number
of non-zero voxels, and thus are well characterized through a sparsity
criterion, even in the case of ICA \cite{varoquaux2010d,daubechies2009}.
However, sophisticated priors such as structured sparsity come with
computational cost and still fail to split some networks into
separate regions.
Altogether, region extraction is unavoidable to go from brain image
decompositions to Regions-of-Interest-based analysis
\cite{Nieto-Castanon2003}.

A simple approach to obtain sharper maps is to use hard
thresholding, which is a good sparse, albeit non convex, recovery method
\cite{blumensath2009}.
We improve upon it by introducing richer post-processing strategies with spatial models,
to avoid small spurious regions and isolate each salient feature in
a dedicated region.
Based on purely geometric properties, these take advantage of the
spatially-structured and sparsity-inducing penalties of recent dictionary
learning methods to isolate regions.
%
%
These can also be used in the framework of computationally cheaper ICA
algorithms.  
In addition to these automatic methods that extract brain atlases,
we propose two metrics to quantitatively compare
them and determine the best one.
The paper is organized as follows. In section 2, we
introduce the region extraction methods. Section 3 presents the
experiments run to compare them.  Finally, results are presented
in section~4.

\section{Region extraction methods}

\begin{figure}[b]
    \begin{center}
    \includegraphics[width=\textwidth]{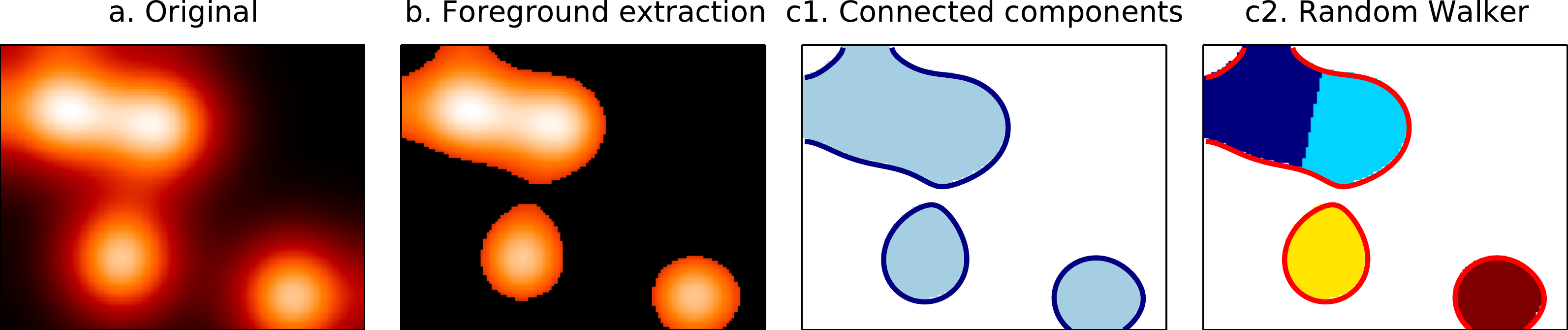}
    \end{center}
    \caption{Example of region extraction. Foreground pixels (b) are extracted
        from the original image (a). Regions are extracted using connected
        component extraction (c1) or random walker (c2).}
\label{fig:cc}
\end{figure}

Extracting regions to outline objects is a well-known problem in computer
vision. 
For the
particular problem of extracting regions of interest (ROIs) out of brain maps, we want a method that
\textit{i)} handles 3D images \textit{ii)} processes one image while
taking into account the remainder of the atlas (e.g., region extraction for a given
image may be different depending on the number of other regions) and
\textit{iii)} isolates each salient feature from a smooth image in an individual
ROI without strong edges or structure (see figure~\ref{fig:cc}).
Here, we assume that a given set of \textit{brain maps} has
been obtained by a multivariate decomposition technique.

%
%

Most of the following methods allow overlapping components after
region extraction.  
In fact, multivariate decomposition techniques most often decompose the
signal of one voxel as a linear mixture of several signal
components.
In practice, these overlapping regions are small and located
in areas of low confidence.  Voxels that belong to no component are
left unlabeled.

\subsection{Foreground extraction}

Let $\mathcal{I} = \{I_1, ..., I_k\}$ be a set of brain maps (3D images), or atlas.
$I(p)$ designates the value for image $I$ at point $p$. We define by
$\mathcal{F}(I)$ the set of foreground points of image $I$, i.e., the points that
are eligible for region extraction. We propose two strategies to extract the
foreground.

\paragraph{Hard assignment.}

Hard assignment transforms a set of maps into a brain segmentation
with no overlap between regions. That means that each voxel will be
represented by a unique brain map from the atlas.
This map is the one
that has the highest value for this voxel. The result is a
segmentation from which we can extract connected components. 

$$\mathcal{F}_{hard}(I_i) = \{p \in I_i\;|\; \text{argmax}_{j \in [1, k]}\,I_i(p) =
i\}$$

\paragraph{Automatic thresholding.}

Thresholding is the common approach used to extract ROIs from
ICA. However, the threshold is usually set manually and is different for
each map.
In order to propose an automatic threshold choice, we consider that on
average, an atlas assigns each voxel to one region. For this purpose,
we set the threshold $t^k(\mathcal{I})$ so that the number of nonzero voxels
corresponds to the number of voxels in the brain:
%


$$\mathcal{F}_{automatic}(I_a) = \{p \in I_a, I(p) > t^k(\mathcal{I})\}$$


\subsection{Component extraction}

\paragraph{Connected components.}

Let $\mathcal{N}(p)$ be the set of neighbors of point $p$.
Two points $p_1$ and $p_n$ are $\mathcal{N}$-connected if $p_n$ can be reached
from $p_1$ by following a path of consecutive neighboring points:
$$(p_1, p_n)\;\mathcal{N}\text{-connected} \equiv \exists (p_2, ...,
p_{n-1}) : p_{i+1} \in \mathcal{N}(p_i), \forall\ i \in [1, n-1]$$
We define a connected component as a maximum set
of foreground points that are $\mathcal{N}$-connected. 
The set of all $\mathcal{N}$-connected components for a given image $I$ (see figure
\ref{fig:cc}.c1) is
written $ccs(\mathcal{N}, I)$. Extraction of connected components can be done
after hard assignment or automatic thresholding to obtain ROIs
(figures~\ref{fig:comparison} and \ref{fig:ica}).
In the following methods, we consider the points extracted with automatic thresholding as
foreground ($\mathcal{F} = \mathcal{F}_{automatic}$) and
use more sophisticated priors to extract ROI. 

\paragraph{Hysteresis thresholding.}

Hysteresis thresholding is a two-threshold method where all voxels
with value higher than a given threshold $t_{high}$ are used as seeds
for the regions. Neighboring voxels with values between the high threshold
$t_{high}$ and the low threshold $t_{low}$ are added to these seed regions.
In our setting, the high
threshold can be seen as a minimal activation value over the regions in order
to sort out regions of marginal importance. Each brain map has its own optimal
value but, in practice, cross validation has shown that keeping the 10\%
highest foreground voxels as seeds gives the best results.
The automatic thresholding strategy described above is used to set the
low threshold $t_{low}$.

Conserving connected components that have their maximum value over $t_{high}$ is
done at component extraction:
$$ccs_{hysteresis}(\mathcal{N}, I) = \{c \in ccs(\mathcal{N}, I)\;|\; \text{max}(c) \geq t_{high}\}$$


\paragraph{Random Walker.}

Random Walker is a seed-based segmentation algorithm similar to watershed. 
It calculates, for each point $p$, the probabilities to end up in each
of the seeds by doing a random walk across the image starting from $p$.
The original version of the algorithm \cite{grady2006} was of probabilistic
nature, whereby the
probability to jump to a neighboring point is driven by the gradient
magnitude between them.
After convergence the point is attached to the seed with the highest
probability. 

Random Walker can also be seen as a diffusion process. It is
equivalent to hysteresis thresholding where regions that have grown enough
to share a boundary are not allowed to be merged. 
The probabilities to reach each of the seeds can be computed using 
the laplacian matrix of the graph associated with the map.
Due to space limitations we refer the reader to \cite{grady2006} for the complete
description of the algorithm. We suppose $seed(p)$ returns the
seed \textit{associated} with point $p$. We refine our neighborhood
relationship by considering two points as neighbours only if they are
\textit{associated} to the same seed:
$$\mathcal{N}_{rw}(p) = \{ q \in \mathcal{N}, seed(p) = seed(q) \}
\; ; \; ccs_{rw}(I) = ccs(\mathcal{N}_{rw}, I)$$



Note that, in our setting, a high value in the map means a high
confidence.  So, instead of using the finite difference gradient,
we consider the max of the image minus the lowest voxel. Therefore,
diffusion is facilitated in areas of high confidence and more
difficult elsewhere. We take the local maxima of the smoothed image as seeds
for the algorithm.

\section{Experiments}

Experiments are made on a subset of the publicly available Autism Brain Imaging Database
Exchange\footnote{\url{http://fcon_1000.projects.nitrc.org/indi/abide/}}
dataset. Preprocessing is done with SPM and includes slice timing, realignment,
coregistration to the MNI template and normalization. 
We select 101 subjects suffering from autism spectrum disorders and 93 typical
controls from 4 sites and compute brain atlases
on 10 cross-validation iterations by taking a random half of the
dataset as the train set. We extract
regions from these atlases and quantify their performance on the other half of
the dataset with two metrics.

We investigate two decomposition methods to extract brain maps from
rest\-ing-state fMRI: \textbf{ICA} --independent component analysis--
that yields full brain continuous maps, and \textbf{MSDL}
--multi-subject dictionary learning--, \cite{abraham2013}, that
directly imposes sparsity and structure on the maps thanks to the
joint effect of $\ell_1$ norm and total variation minimization.
Our goal is to compare the effects of region extraction on sparse and non-sparse
sets of maps.

To quantify the usefulness of a set of regions extracted
automatically, we consider metrics that characterize two different
aspects of the segmentation: the ability to explain newly observed data and the
reproducibility of the information extracted, as in the NPAIRS
framework \cite{strother2002}.
We use Explained Variance (EV) to measure how faithful the extracted
regions are to unseen data. Stability with regards to inter-subject
variability is measured using Normalized Mutual Information (NMI) over
models learned on disjoint subsets of subjects.

Following \cite{varoquaux2011}, we extract $k = 42$ maps.
For the metrics to be comparable, we need to apply them on models of similar
complexity, i.e. with the same number of regions. For this purpose, we assume
that there must be on average 2 symmetric regions per map (some of them may
have more, and some of them may have only one inter-hemispheric region). 
We therefore aim at extracting 2$k$ regions, and take the largest
connected components after region extraction. In the end, some maps
may not contribute to the final atlas.


\subsection{Data faithfulness -- Explained variance}

The explained variance measures how much a model accounts for the variance of
the original data. The more variance is explained, the better the model
explains the original data. 
Linear decomposition models original data $y_{orig}$ by decomposing them into two matrices. In
our case, these matrices are brain networks $\mathcal{I}$ and their associated
time series $y_{model}$.
Time series of regions are measured using least square fitting instead of
simple averaging to handle mixed features in region overlaps.
Explained variance of these series is
then computed over the original ones.
$$y_{orig} = \mathcal{I} \times y_{model} + y_{\varepsilon} \; ; \;
\text{EV}(y_{model}) = 1 -
\frac{\text{Var}(y_{\varepsilon})}{\text{Var}(y_{orig})} =
\frac{\text{Var}(y_{orig}) - \text{Var}(y_{model})}{\text{Var}(y_{orig})}$$

\subsection{Stability -- Normalized Mutual Information}

To assess model stability, we rely on Normalized Mutual Information, a
standard clustering similarity score, applied on hard assignments
\cite{vinh2010}: given two hard assignments $U$ and $V$ with marginal
entropy $H(U)$ and $H(V)$ respectively,
$$NMI(U,V)=\frac{H(U) + H(V) - H(U,V)}{\sqrt{H(U)*H(V)}}\; ;
H(X) = -\sum_{i=1}^n p(x_i)\; log\; p(x_i)$$


\section{Results}

\begin{figure}[p]
    \begin{minipage}{.48\textwidth}
        \begin{tikzpicture}[smooth, xscale=0.38, yscale=.20, inner frame sep=0]
             \node[anchor=south west,inner sep=0] at (0,-4)
                {\includegraphics[width=.5\textwidth]{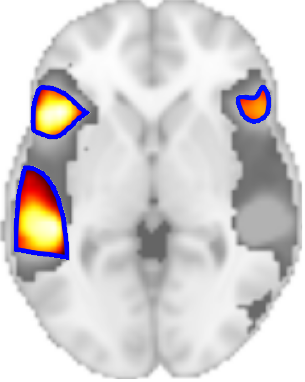}};

            \draw [fill=white] (0, -0.5) rectangle (8, -6.5);
            \begin{scope}
              \clip\imgplot;
              \clip (0,0) rectangle (6, -6);
              \fill[fill=lightgreen] (0,0) -- (0, -4) -- (1.9, -2.8) -- (2.4, -1.6) --
              (3.6, -2.1) -- (4, -4.5) -- (5, -4.7) -- (8, -4.) -- (8, 0) ;
            \end{scope}

            \draw [ultra thick] \imgplot
            \path [ultra thick, red, draw] (0,-4) -- (1.9,-2.8) -- (2.4, -1.6) --
            (3.6, -2.1) -- (4, -4.5) -- (5, -4.7) -- (8, -4.) node {};

        \end{tikzpicture}
        \includegraphics[width=.31\linewidth]{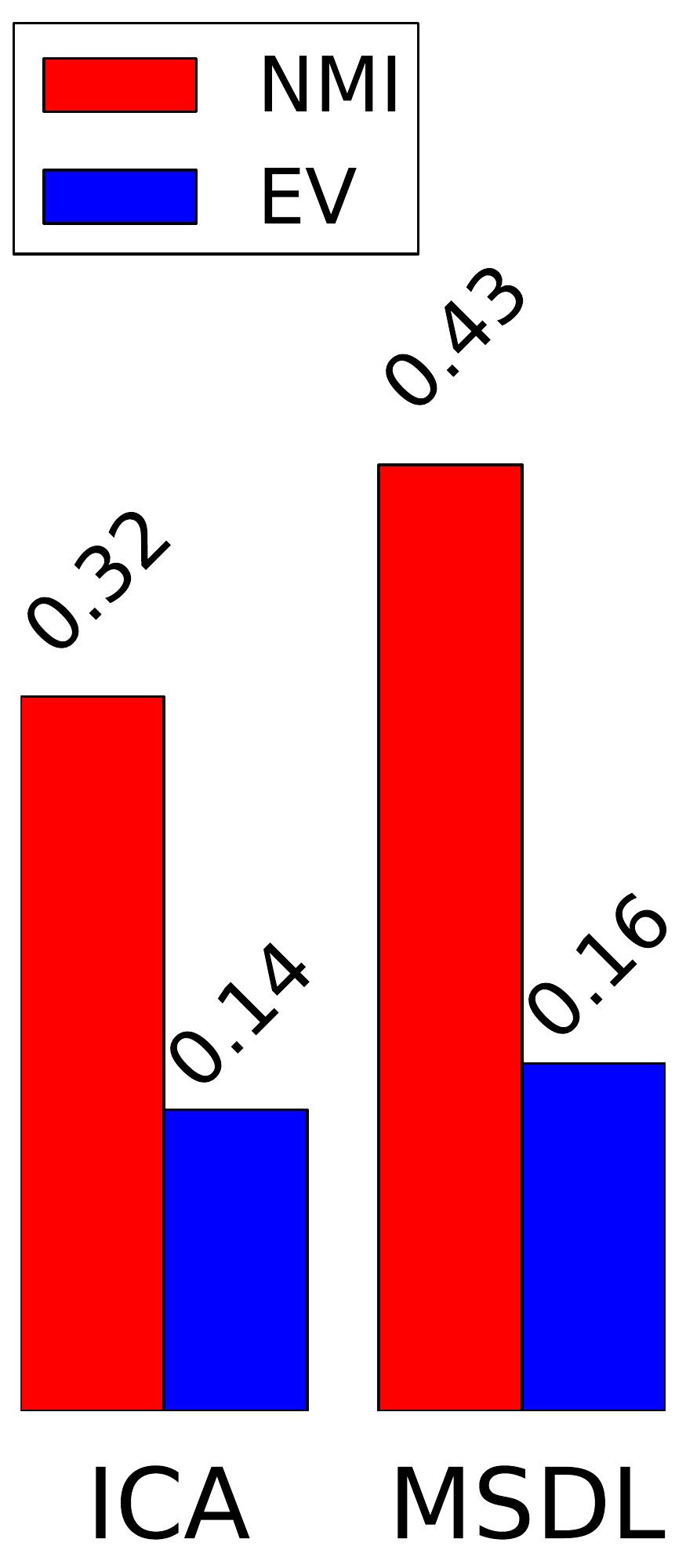}\\
        \centering{\textit{a. Hard thresholding}}
    \end{minipage}
    \begin{minipage}{.48\textwidth}
        \begin{tikzpicture}[smooth, xscale=0.38, yscale=.20, inner frame sep=0]
             \node[anchor=south west,inner sep=0] at (0,-4)
             {\includegraphics[width=.5\textwidth]{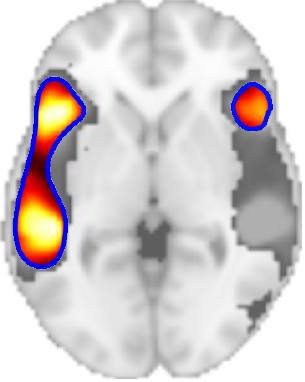}};

            \draw [fill=white] (0, -0.5) rectangle (8, -6.5);
            \begin{scope}
              \clip\imgplot;
              \clip (0,0) rectangle (6, -6);
              \fill[fill=lightgreen] (0,0) rectangle (6,-4.5);
            \end{scope}
            \draw [ultra thick] \imgplot node;
            \draw [ultra thick, red, draw] (0,-4.5) -- (8,-4.5) ;

        \end{tikzpicture}
    \includegraphics[width=.31\linewidth]{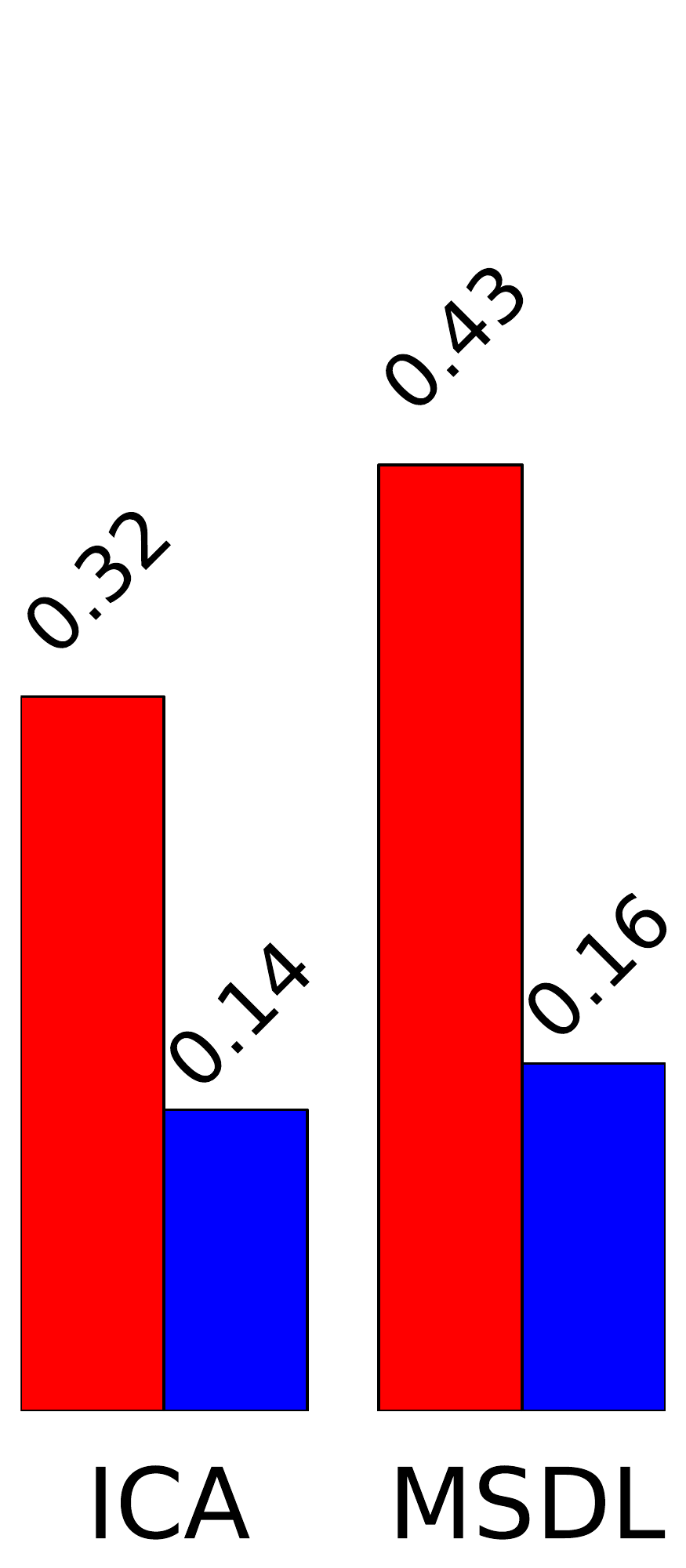}\\
        \centering{\textit{b. Automatic thresholding}}
    \end{minipage}
    \vspace{.2cm}\\
    \begin{minipage}{.48\textwidth}
        \begin{tikzpicture}[smooth, xscale=0.38, yscale=.20, inner frame sep=0]
             \node[anchor=south west,inner sep=0] at (0,-4)
             {\includegraphics[width=.5\textwidth]{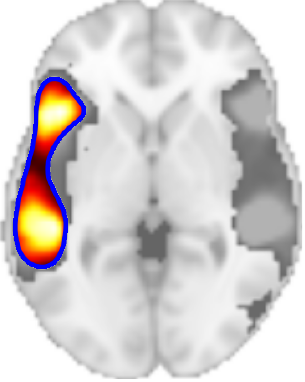}};

            \draw [fill=white] (0, -0.5) rectangle (8, -6.5);
            \begin{scope}
              \clip\imgplot;
              \clip (0,0) rectangle (4, -6);
              \fill[fill=lightgreen] (0,0) rectangle (6,-4.5);
            \end{scope}

            \draw [ultra thick] \imgplot
            \path [ultra thick, blue, draw] (0,-2.5) -- (8,-2.5) node {};
            \path [ultra thick, red, draw] (0,-4.5) -- (8,-4.5) node {};

        \end{tikzpicture}
    \includegraphics[width=.31\linewidth]{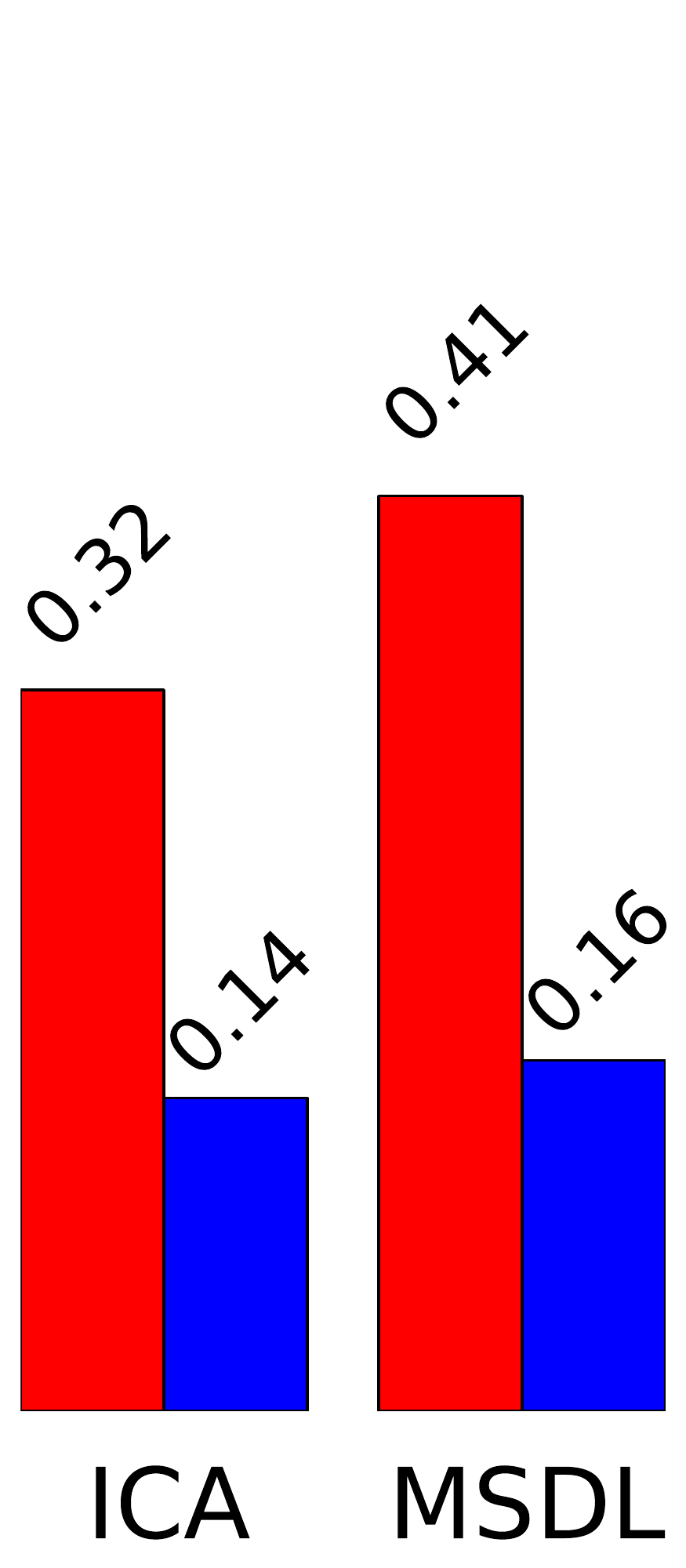}\\
        \centering{\textit{c. Hysteresis thresholding}}
    \end{minipage}
    \begin{minipage}{.48\textwidth}
        \begin{tikzpicture}[smooth, xscale=0.38, yscale=.20, inner frame sep=0]
             \node[anchor=south west,inner sep=0] at (0,-4)
             {\includegraphics[width=.5\textwidth]{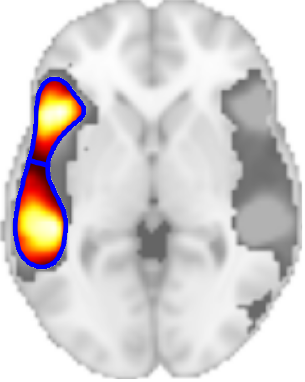}};

            \draw [fill=white] (0, -0.5) rectangle (8, -6.5);
            \begin{scope}
              \clip\imgplot;
              \clip (0,0) rectangle (4, -6);
              \fill[fill=lightblue] (0,0) rectangle (2,-4.5);
              \fill[fill=lightorange] (2,0) rectangle (6,-4.5);
            \end{scope}
    
            \draw [ultra thick] \imgplot
            \path [ultra thick, red, draw] (0,-4.5) -- (7,-4.5) node {};
            \path [ultra thick, blue, draw] (1.1,-1.8) -- (1.1,-2.2);
            \path [ultra thick, blue, draw] (2.9,-0.8) -- (2.9,-1.2);
            \path [ultra thick, red, draw] (1.95,-2.8) -- (1.95,-4.5);

        \end{tikzpicture}
    \includegraphics[width=.31\linewidth]{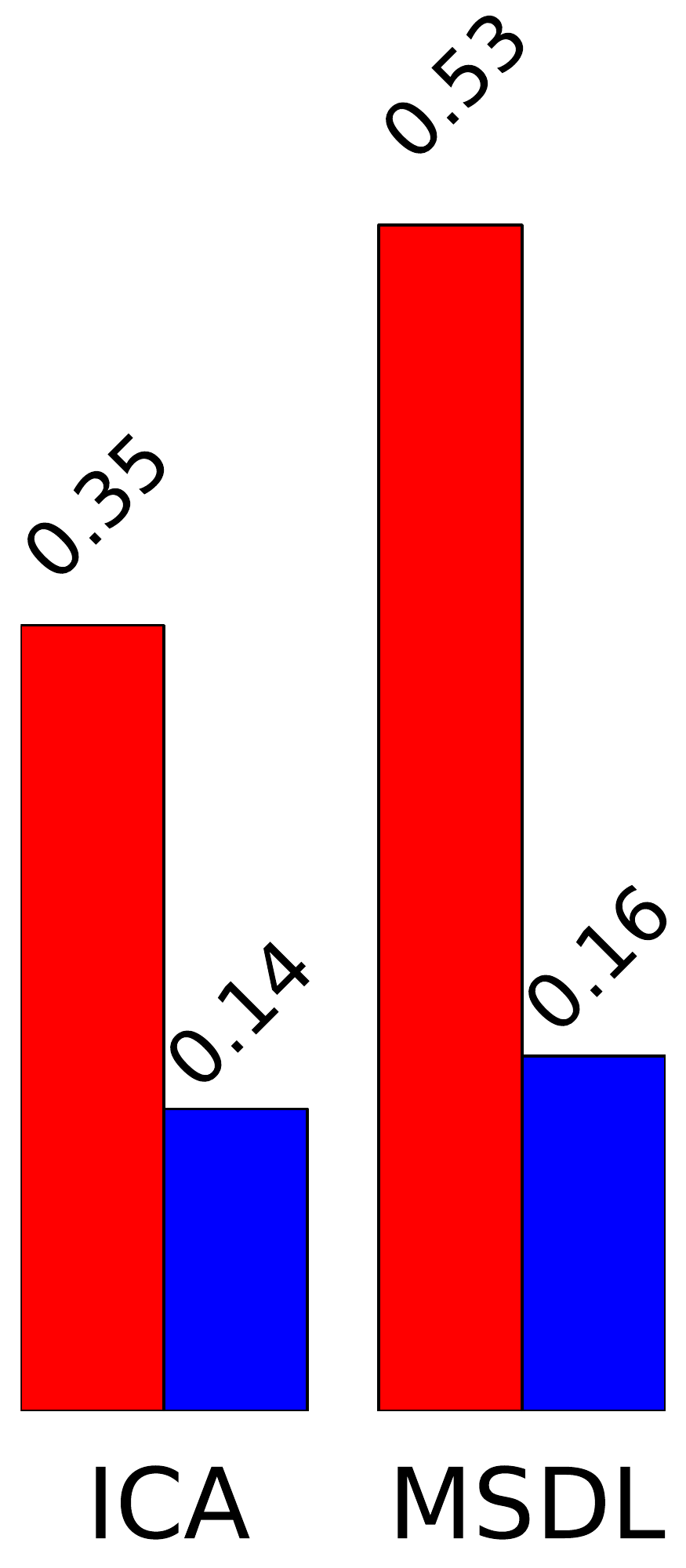}\\
        \centering{\textit{d. Random Walker}}
    \end{minipage}
    \caption{Comparison of region extraction methods (after selection of 2$k$
        regions). Brain maps obtained with
      MSDL are located on the left. The activated regions are symbolically
      represented below in a height map.
      The bars on the right of each image represent the
      Normalized Mutual Information and Explained variance obtained on
      dense maps (ICA) and sparse maps (MSDL). Random walker is the most
  stable method.}
    \label{fig:comparison}
\end{figure}
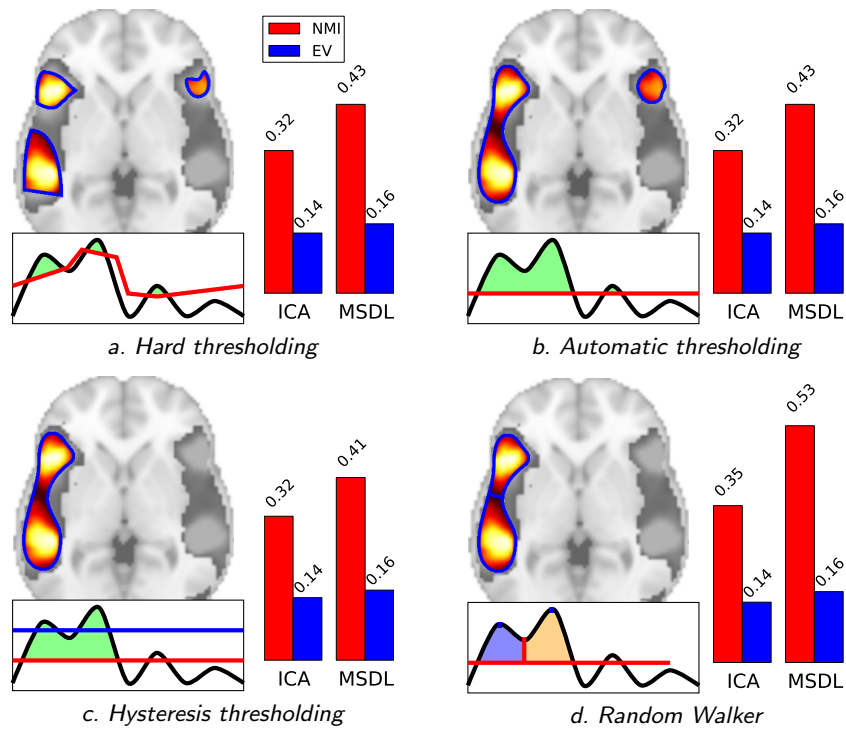

\begin{figure}[p]
    \textbf{Visual cortex}\\[.1cm]
    \begin{tabular}{cccccc}
        \includegraphics[width=.11\linewidth]{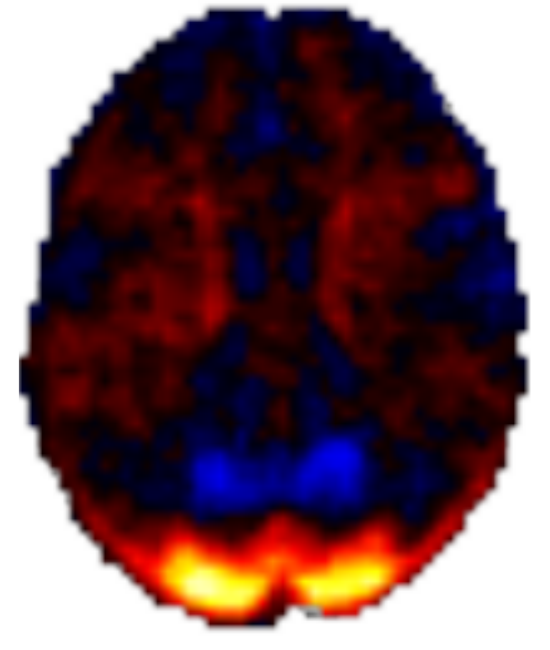} &
        \includegraphics[width=.11\linewidth]{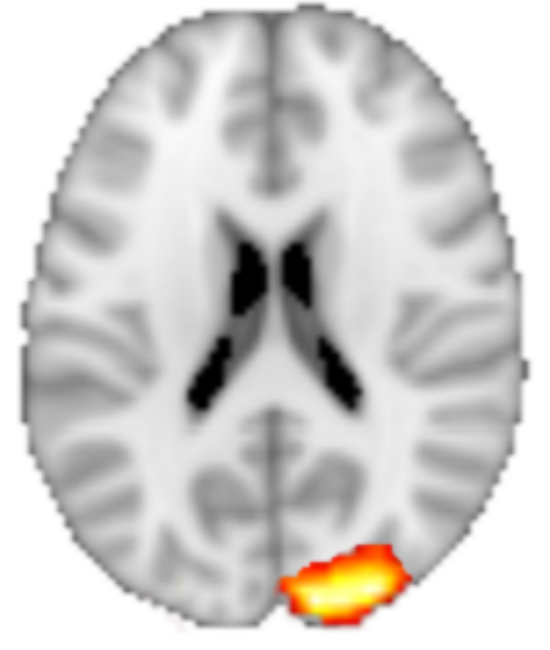}
        \includegraphics[width=.11\linewidth]{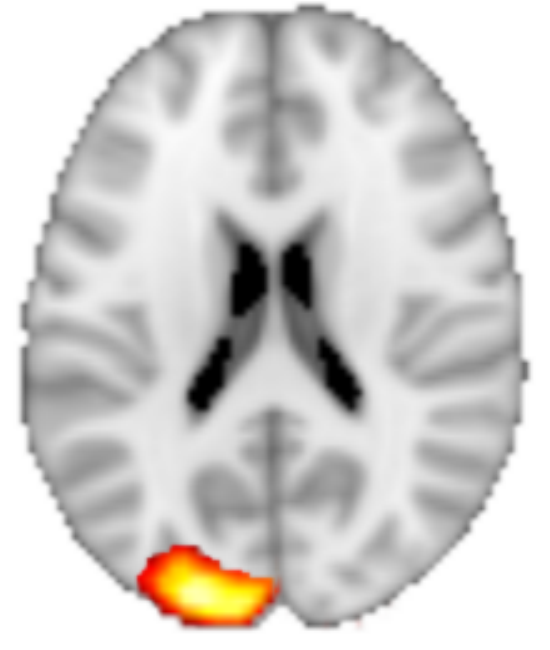} &
        \includegraphics[width=.11\linewidth]{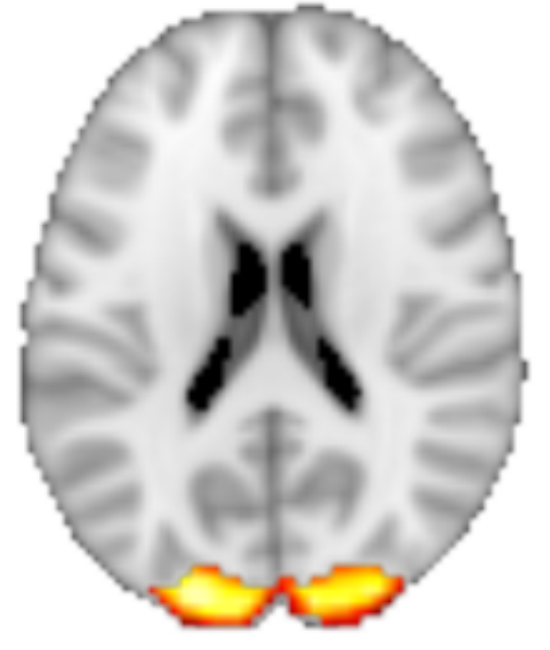} &
        
        \includegraphics[width=.11\linewidth]{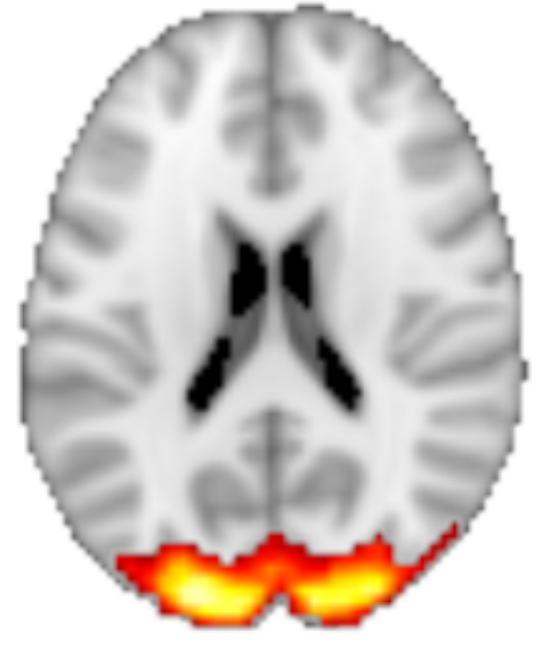} &
        
        \includegraphics[width=.11\linewidth]{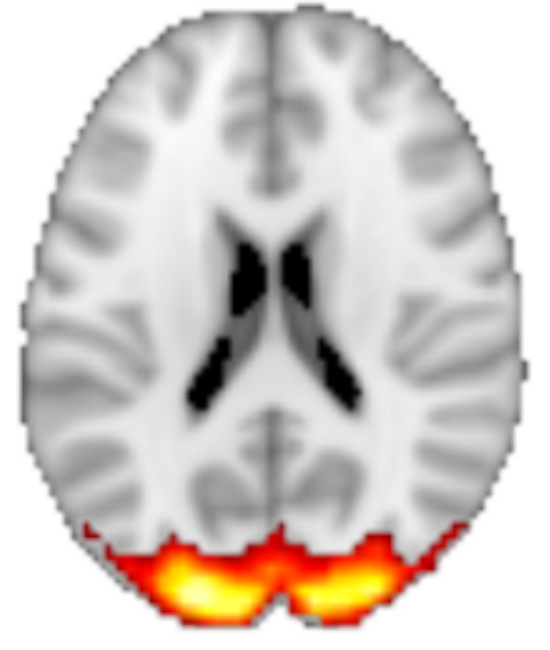} &
        \includegraphics[width=.11\linewidth]{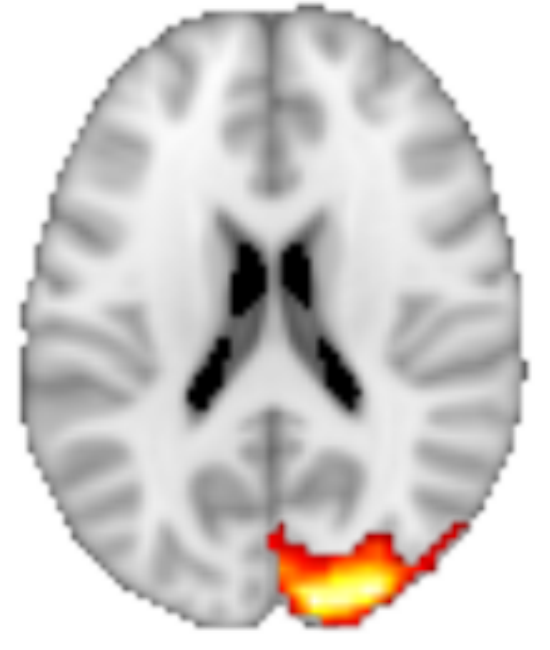}
        \includegraphics[width=.11\linewidth]{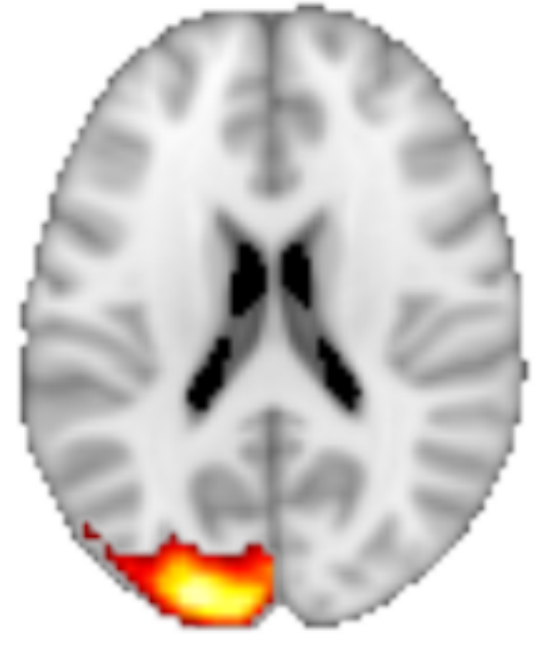}
        \\[-.2cm]
        & \upbracefill & & & & \upbracefill \\
        Original & Manual & Hard & Automatic & Hysteresis & Random Walker\\
    \end{tabular}\\[.5cm]
    \textbf{Default mode network}\\[.1cm]
    \begin{tabular}{ccccc}
        \includegraphics[width=.11\linewidth]{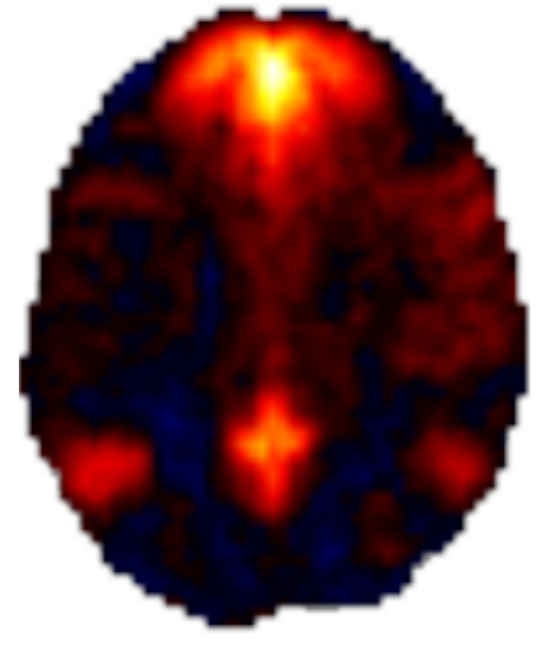} &
        \includegraphics[width=.11\linewidth]{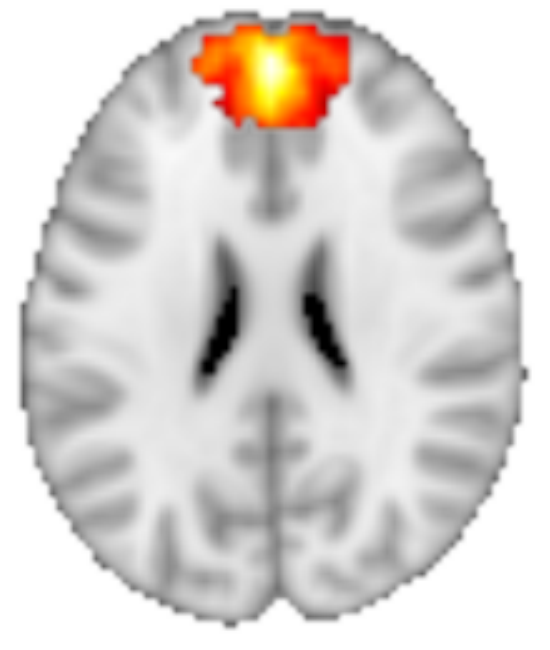} &
        
        \includegraphics[width=.11\linewidth]{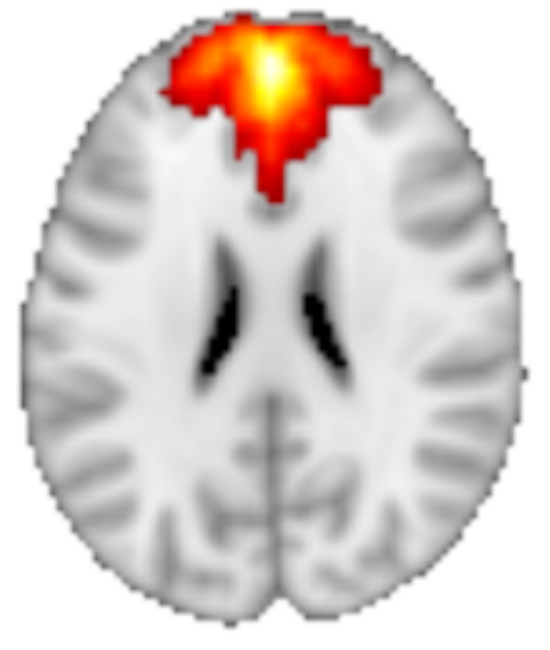}
        \includegraphics[width=.11\linewidth]{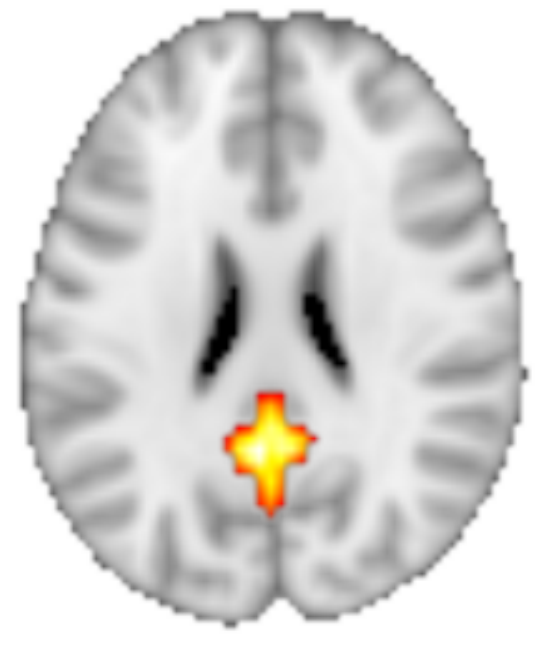} &
        
        \includegraphics[width=.11\linewidth]{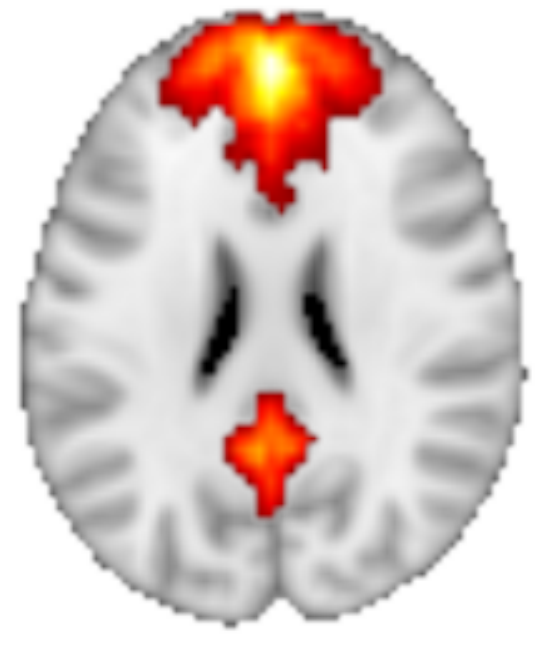} &
        \includegraphics[width=.11\linewidth]{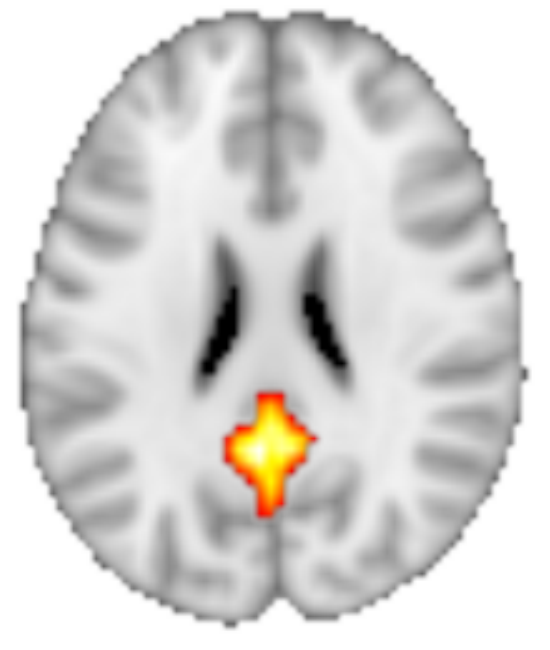}
        \includegraphics[width=.11\linewidth]{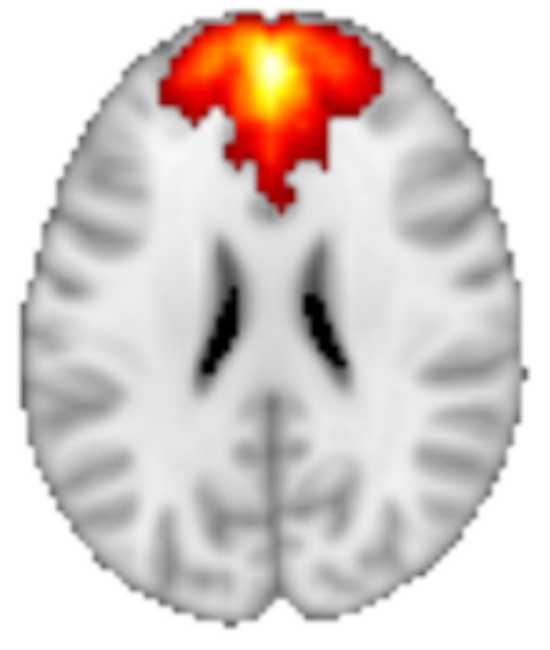}
        \includegraphics[width=.11\linewidth]{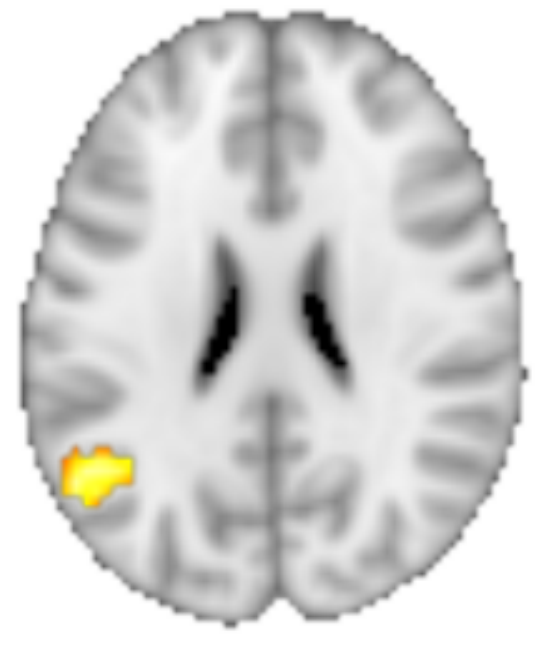}
        \\[-.2cm]
        & & \upbracefill & & \upbracefill \\
        Original & Hard & Automatic & Hysteresis & Random Walker\\
    \end{tabular}\\
    \label{fig:ica}
\end{figure}


Figure~\ref{fig:comparison} presents region extraction results using
each method on the same map.
In all figures, the threshold applied during region extraction
is shown in a given slice to help understanding. 
Results for each metric are displayed on the right. We vary parameters
 for each model (smoothing for ICA, 3 parameters of MSDL) and, for each
region extraction method, display the best 10\% results across
parametrization.
Figure~\ref{fig:ica} shows 2 networks out of 42 extracted.

\paragraph{Region shape} The regions extracted by hard assignment
(figure~\ref{fig:comparison}.a) present salient angles and their
limits do not follow a contour line of the original map. The straight lines
are the results of two maps in competition with each other. The 1D cut shows 
that the threshold applied when using hard thresholding is not uniform
on the whole image.
The other methods look smoother and follow actual contour lines of
the original map.
On this particular example, automatic thresholding
(figure~\ref{fig:comparison}.b) extracts 2 regions: a large one on the left
and a very small one on the right. 
This is one of the drawbacks of thresholding: small regions can appear
when their highest value is right above the threshold.
Thanks to its high threshold, hysteresis thresholding
(figure~\ref{fig:comparison}.c) gets rid of the spurious regions but
still fails to separate the large region on the left.  Random Walker
(figure~\ref{fig:comparison}.d) manages to split the large region into
two subregions.

Similarly, in figure~\ref{fig:ica} we can see that
Random Walker manages to split the default mode network into 3 components,
where other methods extract two.

\paragraph{Stability.} Random Walker dominates the stability metric.
It uses local maxima to get regions seeds, and will thus
split regions even if they are \textit{connected} after thresholding.
Its performance is statistically significant for both dense and sparse
atlases and any parametrization.
The stability improvement is larger for sparse than for
dense maps. This could be due to the inability of random walker to compensate
for the original instabilities of the models.

\paragraph{Data fidelity.} The explained variance scores on best performing models, shown in figure~\ref{fig:comparison}, are similar for all
methods. In poorly performing models, we observe that automatic and
hysteresis thresholdings are slightly above random walker (about 2\%),
exhibiting the same trade-off as in \cite{abraham2013}.

\section{Discussion and conclusion}

Functional atlases extracted using ICA or sparse decomposition methods are
composed of continuous maps and sometimes fail to separate
symmetric functional regions.
%

Starting from hard thresholding \cite{blumensath2009},
we introduce richer strategies integrating spatial models,
to avoid small spurious regions and isolate each salient feature in
a dedicated region.
Indeed, the notion of regions is hard to express with convex penalties.
Relaxations such as total-variation used in \cite{abraham2013} only
captures it partially, while a non-convex segmentation step easily
enforces regions.
We find that a Random-Walker based strategy brings substantial increase
in stability of the regions extracted, while keeping very good
explanatory power on unseen data. Finer results and interpretation may arise by
using more adapted metrics, for example a version of DICE that can deal with
overlapping fuzzy regions. This point is under investigation.

\noindent\textbf{Acknowledgments} We acknowledge funding from the NiConnect
project and NIDA R21 DA034954, SUBSample project from the DIGITEO
Institute, France.


\bibliography{bibliostmi}

%
%
%
%

\end{document}